%% file: main.tex
\documentclass[a4paper]{article}

\usepackage{INTERSPEECH2022}

\title{On the Role of Spatial, Spectral, and Temporal Processing for DNN-based Non-linear Multi-channel Speech Enhancement}
\name{Kristina Tesch, Nils-Hendrik Mohrmann, and Timo Gerkmann}
\address{
  Signal Processing, Universität Hamburg, Germany}
\email{firstname.lastname@uni-hamburg.de}

\usepackage{microtype}
\usepackage[nolist]{acronym}
\usepackage{dirtytalk}
\usepackage{booktabs, dcolumn}
\usepackage{gensymb}
\usepackage{url}
\usepackage[moderate]{savetrees}
\usepackage{cite}
\usepackage[inline]{enumitem}
\usepackage{pifont}
\usepackage{placeins}
\usepackage[hidelinks]{hyperref}
\usepackage[font=small,skip=3pt]{caption}
\usepackage[keeplastbox]{flushend}

\usepackage{physics, amsmath}
\usepackage{IEEEtrantools}

\usepackage{caption}
\usepackage{subcaption}
\usepackage[export]{adjustbox}
\usepackage{graphicx}
\usepackage{pgfplots}
\DeclareUnicodeCharacter{2212}{−}
\usepgfplotslibrary{groupplots,dateplot}
\usetikzlibrary{quotes,arrows.meta, calc, patterns.meta, shapes.multipart, shapes.arrows, shapes.misc, spy, decorations.pathreplacing, arrows.meta, backgrounds, shapes.geometric, intersections}
\pgfplotsset{compat=1.15}

\input{acronyms}

\input{notation}

\begin{document}

\maketitle
\begin{abstract}
Employing \acp{DNN} to directly learn filters for multi-channel speech enhancement has potentially two key advantages over a traditional approach combining a linear spatial filter with an independent tempo-spectral post-filter: 
\begin{enumerate*}[label=\arabic*)]
    \item non-linear spatial filtering allows to overcome potential restrictions originating from a linear processing model and 
    \item joint processing of spatial and tempo-spectral information allows to exploit interdependencies between different sources of information.
\end{enumerate*}

A variety of DNN-based non-linear filters have been proposed recently, for which good enhancement performance is reported. However, little is known about the internal mechanisms which turns network architecture design into a game of chance. Therefore, in this paper, we perform experiments to better understand the internal processing of spatial, spectral and temporal information by DNN-based non-linear filters. 

On the one hand, our experiments in a difficult speech extraction scenario confirm the importance of non-linear spatial filtering, which outperforms an oracle linear spatial filter by 0.24 POLQA score. On the other hand, we demonstrate that joint processing results in a large performance gap of 0.4 POLQA score between network architectures exploiting spectral versus temporal information besides spatial information. 
\end{abstract}

\section{Introduction}
{\let\thefootnote\relax\footnote{\footnotesize We thank Rohde\&Schwarz SwissQual AG for support with POLQA.}}
Speech enhancement algorithms are employed to improve the speech quality and speech intelligibility of speech signals recorded in noisy and often reverberant environments. Their use is indispensable for many applications that are required to work reliably in unfavorable acoustic scenarios, e.g., automatic speech recognition or hearing aids. Accordingly, research on this topic has been ongoing for decades. 

Many algorithms operate in the \ac{STFT} domain and traditionally rely on a statistical model to derive an analytical clean speech estimator, e.g., \cite{1984ephraimSpeechEnhancementUsing, lotter2005SpeechEnhancementMAP, Erkelens2007MinimumME, becker2016}. However, simplifying assumptions must often be made to keep the problem tractable. For example, neighboring time-frequency-bins are often assumed to be independent. In contrast, recent state-of-the-art single-channel speech enhancement algorithms are built from \acp{DNN} \cite{tan2018convolutional, giri2019waveUnet, koizumi2021sepformer}, which do not require an explicit model but learn complex dependencies directly from data. It is common knowledge that correlations in the time and the frequency dimension should be exploited by \acp{DNN} for good performance \cite{richter2020speech, tang2021joint}. 

\begin{figure}
    \begin{minipage}[]{0.1\linewidth}
        \subcaption{}\label{fig:1-separated}
    \end{minipage}
    \begin{adjustbox}{minipage=0.85\linewidth}
        \centering
        \includegraphics{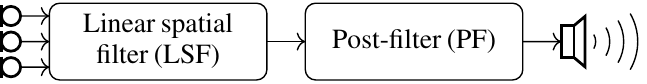}
    \end{adjustbox}
    
    \begin{minipage}[]{0.1\linewidth}
        \subcaption{}\label{fig:1-jointfilter}
    \end{minipage}
    \begin{adjustbox}{minipage=0.85\linewidth}
        \centering
        \includegraphics{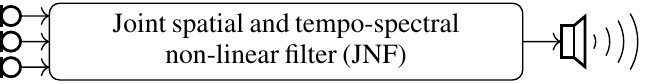}
    \end{adjustbox}
    
    \begin{minipage}[]{0.1\linewidth}
        \subcaption{}\label{fig:1-separated2}
    \end{minipage}
    \begin{adjustbox}{minipage=0.85\linewidth}
        \centering
        \includegraphics{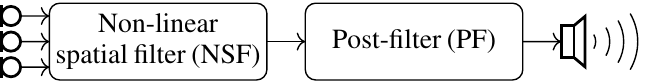}
    \end{adjustbox}
    \caption{(a) The traditional two-step processing using a linear spatial filter (beamformer) followed by a single-channel post-filter. (b) A joint spatial and tempo-spectral non-linear processing scheme that we implement using \acp{DNN} in this work. (c) Two-step processing scheme, however, not only the post-filter performs non-linear filtering but also the spatial filter.}
    \label{fig:1-comparison}
\end{figure}

If the noisy signals are recorded with multiple microphones, then spatial information is available in addition to tempo-spectral information. Traditional approaches usually follow the two-step approach illustrated in Figure \ref{fig:1-separated}, which first applies a linear spatial filter, a so-called beamformer \cite[Sec. 12.4.2]{vary2006digital}, and then employs a single-channel post-filter to exploit tempo-spectral information \cite{Simmer2001, doclo2015MultichannelSignalEnhancement}. While such a separated setup is often not considered a limitation, in our prior work \cite{tesch2019NonlinearSpatialFiltering, tesch2021tasl}, we have demonstrated that just assuming non-Gaussian distributed noise leads to a \ac{MMSE} estimator that combines spatial and spectral processing into a single non-linear operation, which has superior performance over a linear spatial filter combined with a post-filter. While experiments with the analytic estimator show great potential for joint non-linear spatial-spectral processing, practical applicability is questionable because accurate parameter estimation of higher-order statistics required the use of oracle knowledge. However, \acp{DNN} provide a data-driven way to implement practical joint spatial and tempo-spectral non-linear filters (JNF). See Figure \ref{fig:1-jointfilter} for an illustration.

While DNN-based approaches have been dominating the single-channel speech enhancement research for a couple of years now, many publications on multi-channel speech enhancement have proposed to combine \acp{DNN} with traditional methods, e.g., \cite{heymann2016nnestimation, togami2019itakura}. However, the potentially greatest advantage of using DNNs, allowing for non-linear instead of linear spatial processing and taking the interdependencies between spatial and tempo-spectral processing into account, cannot be exploited this way. This is different for the variety of data-driven multi-channel filters that have been proposed recently \cite{li2019narrowband, chakrabarty2019, tolooshams2020cadunet, wang2020complexspectralmapping}. These approaches report good performance for speech enhancement tasks, but their internal mechanisms are not well understood. However, this is essential for a deliberate design of a network architecture that fully unlocks the potential of neural networks for multi-channel speech enhancement. 

\begin{figure*}[htb]
    \centering
    \includegraphics{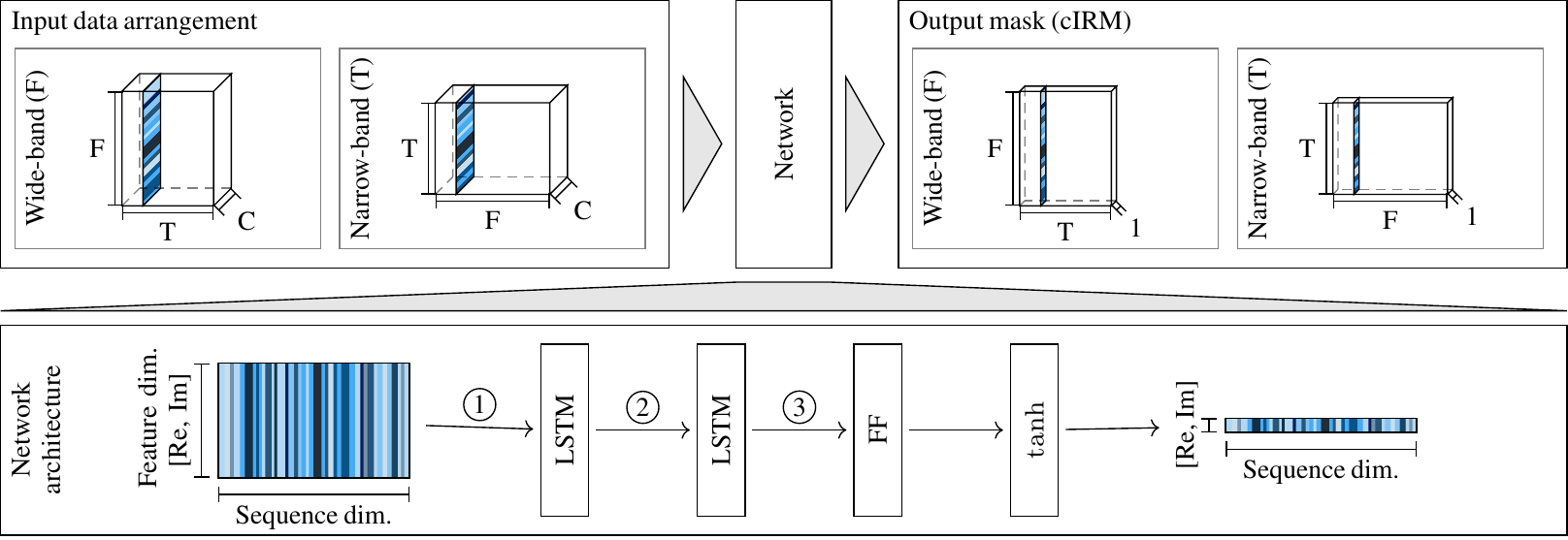}
    \caption{Illustration of the base system architecture. The input data is arranged according to a wide-band or narrow-band input and fed into a network with two LSTM layers, an FF layer and $\tanh$ activation to obtain an estimate of a cIRM. }
    \label{fig:base-arc}
    \vspace{-2em}
\end{figure*}

In this work, we investigate this internal functioning of DNN-based non-linear filters for multi-channel speech enhancement. We aim to answer the following research questions: Is non-linear as opposed to linear spatial filtering the main factor for good performance? Or is it rather the interdependency between spatial and tempo-spectral processing? In the first case, we could independently perform the non-linear spatial and tempo-spectral processing as shown in Figure \ref{fig:1-separated2}, which would be advantageous for practical applications as this allows for independent optimization of either part. Or, on the contrary, is the consideration of interdependencies between spatial and tempo-spectral information particularly important? And do temporal and spectral information have the same impact on spatial filtering performance? 
\FloatBarrier

We experimentally address these research questions using a set of DNN-based filter variants derived from a simple base network architecture (outlined in Section \ref{ss:netarc}), which are then applied to a challenging speech extraction scenario (described in Section \ref{ss:data}) that provides a good sense of the spatial filtering capabilities of the employed networks. We provide results for experiments on the separability of spatial and tempo-spectral processing in Section \ref{sec:separability} and on the contribution of different information sources in Section \ref{sec:contribution}.

\section{DNN-based non-linear filtering for multi-channel speech enhancement}\label{sec:dnn-based-filtering}

We consider a target clean time-domain signal $s(t)$, which is recorded by a microphone array with $C$ microphone channels in a noisy reverberant room. The recording of $s(t)$ by the $\ell$'s microphone $x^{(\ell)}(t)$ will then not only undergo a time-shift due to the propagation delay from the target source position to the microphones but also include reverberation \cite{doclo2015MultichannelSignalEnhancement}. Transforming to the STFT domain leads to the complex-valued coefficient $X^{(\ell)}(k,i)$ with frequency-bin index $k$ and time-frame index $i$. We use a bold symbol to denote a vector stacking all microphone signals, e.g., $\vecX(k,i) = [X^{(0)}(k,i), ...,X^{(C-1)}(k,i)] \in\mathbb{C}^M$, and drop the indices $(k,i)$ to denote the (multi-channel) spectrogram, e.g., $\vecX\in\mathbb{C}^{C\times F\times T}$ with $F$ and $T$ denoting the number of frequency-bins and time-frames respectively. We assume that speech $\vecX$ and noise signal $\vecV$ sum at the microphones to obtain the noisy signal
\begin{equation}
    \vecY(k,i) = \vecX(k,i)+\vecV(k,i).
\end{equation}
In our experiments, we use multiple interfering speech sources as noise signal. Similar to the target speech signal, also the interfering signals recorded at the microphones will incorporate spatial information related to the positioning of sources and the characteristics of the room. Given the noisy recording $\vecY\in\mathbb{C}^{C\times F\times T}$, we aim to recover the clean target speech signal $S\in\mathbb{C}^{F\times T}$ except for a time-shift caused by the propagation delay to the reference microphone, for which we pick the first channel. 

\subsection{Network architectures}\label{ss:netarc}
The focus of this work is to investigate DNN-based non-linear spatial filters for multi-channel speech enhancement in order to better understand the contribution of individual sources of information (spatial, spectral, and temporal) as well as their interdependencies. For this, we develop a number of DNN-based multi-channel filters derived from an \ac{LSTM} network architecture, which has been proposed by Li and Horaud \cite{li2019narrowband}. 

\subsubsection{Base LSTM architecture (F-JNF, T-JNF)}
The base architecture is depicted in Figure \ref{fig:base-arc}. The multi-channel input (top left) is fed into a neural network (bottom) to obtain a compressed estimate ($C=K=1$ as defined in \cite{williamson2016cirm}) of the target speech \ac{cIRM} $\mathcal{M}_S(k,i)\in\mathbb{C}$ (top right). The target speech signal estimate $\hat{S}(k,i)\in\mathbb{C}$ for every time-frequency-bin $(k,i)$ is then obtained by multiplication of the uncompressed estimated speech mask with the reference channel's noisy recording $Y^{(0)}(k,i)$, i.e., 
\begin{equation}\label{eq:speechest}
    \hat{S}(k,i) = \mathcal{M}_S(k,i) ~ Y^{(0)}(k,i).
\end{equation}
The network architecture is deliberately kept simple with only two bi-\ac{LSTM} layers followed by a feed forward layer and a $\tanh$ activation. As standard \ac{LSTM} layers can only process two-dimensional data (a sequence of features), slices of the three-dimensional input are processed independently with the real and imaginary parts being stacked in the channel (feature) dimension. In their work, Li and Horaud \cite{li2019narrowband} propose to independently process the time-sequence of STFT coefficients $\vecY(k, \cdot)\in\mathbb{C}^{C\times T}$ for all frequency-bins $k=0,...,F-1$. In Figure \ref{fig:base-arc}, this is illustrated as narrow-band input data arrangement and throughout this work we will refer to this as the temporal information based joint non-linear filter (T-JNF). This filter can utilize the fine-grained spatial information in the channel dimension and temporal information along the time axis, however, it does not have access to fine-grained spectral information. In this work, we propose a superior wideband processing scheme (F-JNF) that processes every time-step independently but combines fine-grained spatial and spectral information for mask estimation. 

\subsubsection{Combining temporal and spectral information (FT-JNF)}\label{subs:combined}
While the base architecture combines spatial information with either spectral or temporal information, we now combine all three sources of information within the same general network architecture (leaving the number of parameters unchanged). For this, we propose to simply switch the the data arrangement from wide-band to narrow-band between the two \ac{LSTM} layers at the position marked with \ding{193}. 
This way, information in all three dimensions can be exploited and the filter is denoted as FT-JNF.

\subsubsection{Non-linear spatial filtering (T-NSF, F-NSF, FT-NSF)}
To investigate the non-linear spatial filtering capabilities of the DNN-based filter, we adapt the base network architecture to exclude the second source of fine-grained temporal or spectral information. This is done by randomly shuffling every input along the sequence dimension before feeding the input to the first \ac{LSTM} at the position marked with \ding{192}. This way, only global statistics along the sequence dimension are accessible but correlations between neighboring sequence elements cannot be exploited. However, in a wide-band setup (F-NSF), we noticed that the networks have problems of exploiting spatial information if the frequency bin index is unknown to the network. The frequency-bin index is likely a very important source of information for spatial processing as spatial characteristics strongly depend on the frequency. For this reason, we append the frequency-bin index to the channel dimension such that this information is still available after shuffling along the sequence dimension. For better comparability, we also add the frequency-bin index to the narrow-band setup (T-NSF), however, this has only a small impact on the performance. The permutation of the sequence is then undone after the two LSTM layers (\ding{194}). As in Section \ref{subs:combined}, we define FT-NSF, which switches from narrow-band to wide-band data arrangement at position \ding{193}, however, requiring both LSTM layers to be wrapped in permutation and inverse permutation steps. 

\subsubsection{DNN-based post-filtering (PF)}\label{ss:na:pf}
Last, we train a single-channel post-filtering scheme based on the \ac{LSTM} network architecture denoted by PF. In the single-channel setup, the input has only two dimensions (time and frequency). Here we stack the real and imaginary parts in the frequency dimension, which will serve as feature dimension, while the time dimension corresponds to the sequence dimension. 

\subsection{Data simulation}\label{ss:data}
For our analyses, we generate a simulated dataset with six spatially displaced speech sources of which one target source is to be extracted. As the target and interfering signal (five speakers) have similar tempo-spectral characteristics, the target speaker has to be identified by its spatial location. Accordingly, we place the target source in a fixed angle with respect to the microphone orientation. 
\begin{figure}[tb]
\begin{tabular}{l l}
\parbox{0.55\linewidth}{
\includegraphics{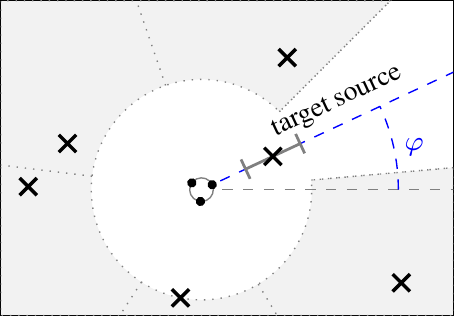}
}
&
\parbox{0.35\linewidth}{
\footnotesize
\begin{tabular}{lr}\toprule
        \multicolumn{2}{c}{\normalsize Room characteristics}\\\midrule
         Width & $2.5-5$ m\\
         Length & $3-9$ m\\
         Height & $2.2-3.5$ m\\
         T60 & $0.2 - 0.5$ s\\\bottomrule
    \end{tabular}
}
\end{tabular}
    \caption{Illustration of the simulation setup. The target source is located in a fixed orientation with respect to microphone array. The five interfering sources are placed in the gray area (one per segment). Room properties are sampled from the given ranges.}
    \label{fig:simu}
\end{figure}

An illustration of the setup simulated with pyroomacoustics \cite{scheibler2018} is depicted in Figure \ref{fig:simu}. For each sample, the room dimensions and reverberation time are uniformly sampled from the ranges listed on the right. The uniform circular microphone array has three channels and a diameter of $10$~cm. Its position in the xy-plane is sampled to have a minimum distance of $1$~m to the walls and placed at height $1.5$~m. Furthermore, we randomly rotate the microphone array. The rotation $\varphi\in [0,2\pi)$ is indicated by the dashed blue line in Figure \ref{fig:simu}. The target source is placed on the blue line with a minimum distance of $0.3$~m and up to $1$~m away from the microphone array. Five interfering sources are placed in the gray area leaving a $1$~m distance to the microphone array and 20\degree~to the position of the target source with one interfering speech source per segment as indicated by the dotted gray lines. The height of the interfering speech sources is sampled from a normal distribution with mean $1.6$~m and standard deviation $0.08$.

We generate 6000, 1000, and 600 samples with a sampling frequency of $16$~kHz for training, validation and testing respectively using clean speech signals from the WSJ0 dataset \cite{wsjdata2007}. Signals between the different sets do not overlap. The \ac{SNR} is not explicitly controlled but obtained from the the simulation setup with varying distances of the sources to the microphone array. The average SNR is $-4$~dB and $95$\% of the data samples distribute between $-9$~dB and $2$~dB.

\subsection{Training details}
For training the multi-channel networks (all except PF), we have access to the noisy observations $\vecy(t)$, the noise signals $\vecv(t)$ and the dry signal $s(t)$, which has been aligned with the noisy observation to include the propagation path delay. We randomly extract three seconds of audio from the utterances in each training iteration and compute the \ac{STFT} using a window length of $32$ ms and $50\%$ overlap with a $\sqrt{\text{Hann}}$ window for analysis and synthesis. 
Using the relationship between the real part $\Re(\cdot)$ and the imaginary part $\Im(\cdot)$ of the speech and noise mask
\begin{IEEEeqnarray}{rClCrCl} 
\Re(\mathcal{M}_\text{V}) &=& 1 - \Re(\mathcal{M}_\text{S}),&\quad&\Im(\mathcal{M}_\text{V})&=&-\Im(\mathcal{M}_\text{S}),
\end{IEEEeqnarray}
we obtain an estimate of the noise mask $\mathcal{M}_\text{V}$. From this, an estimate of the noise signal $\hat{V}$ is computed by applying the noise mask in an analog way as the speech mask (see (\ref{eq:speechest})). We use the loss function proposed by Tolooshams et al. \cite{tolooshams2020cadunet}, which is composed of time and frequency domain $\ell_1$ loss terms:
\begin{equation}
    L(u, \hat{u}) = \sum_{u\in\{s, v\}} \alpha\norm{u-\hat{u}}_1 + \norm{|U|-|\hat{U}|}_1. 
\end{equation}
We set $\alpha=10$ to equalize the contribution of either domain in the loss term. If the ground truth for the noise signal is unknown, we only use the clean speech related parts of the loss function. 

We train the networks with batch size six until convergence (max. 250 epochs) and select the best model based on the validation loss. The number of LSTM units is set to 256 and 128 for all networks, except PF, for which 256 units are used in both layers. 

\section{Separability of spatial processing and post-filtering}\label{sec:separability}

\begin{figure}[t]
    \centering
    \includegraphics{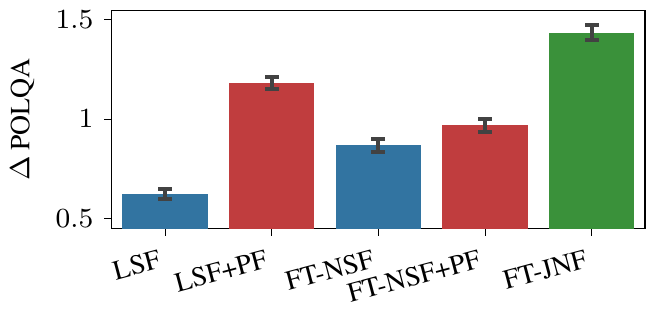}
    \vspace{-0.6em}
    \caption{A comparison of approaches combining a spatial filter (blue) with a post-filter (red) and a joint approach (green). The bars show mean POLQA scores and the 95\% confidence interval.
    }
    \label{fig:separated-processing}
\end{figure}

\begin{figure}[t]
\vspace{-1.4em}
    \centering
    \includegraphics{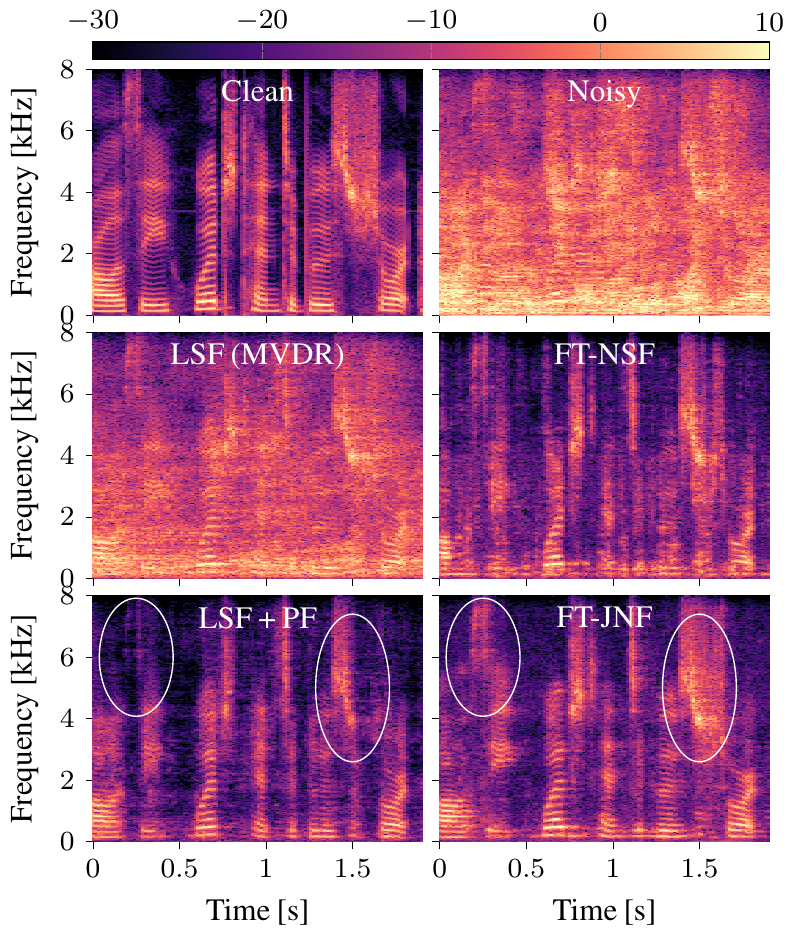}
\vspace{-1.2em}
    \caption{Spectrogram visualization of an example utterance.}
\vspace{-0.5em}
    \label{fig:spectrograms}
\end{figure}

Using the \ac{DNN}-based filters outlined in the previous section, we investigate if multi-channel non-linear filtering can be separated into spatial processing and single-channel post-filtering. For this, we compare the performance of the three approaches illustrated in Figure \ref{fig:1-comparison}. The mean POLQA improvement scores \cite{polqa2018} along with the $95$\% confidence interval are presented in Figure \ref{fig:separated-processing}. 
The POLQA algorithm provides a measure of speech quality on a \ac{MOS} scale ranging from 1 (low quality) to 5 (high quality). The blue bars in Figure \ref{fig:separated-processing} correspond to spatial-only filters. We compute the traditional linear \ac{MVDR} \cite{vary2006digital} based on oracle parameter estimates. A time-varying noise covariance estimate is obtained via recursive averaging of the oracle data and the acoustic transfer function (ATF) is estimated by multiplying the principal eigenvector of the generalized eigenvalue problem for speech and noise covariance matrices with the speech covariance matrix \cite{ito2017}. As the parameters are very accurately estimated from oracle data, the displayed results achieved by the MVDR should be considered as an upper bound for the performance that is achievable with the linear processing model. Nevertheless, the oracle MVDR is outperformed by the DNN-based non-linear spatial filter (FT-NSF) evaluated on unseen test data by 0.24 POLQA score. The differences between the two estimates are clearly visible in the middle row of Figure \ref{fig:spectrograms}. While the MVDR obeys a distortionless constraint at the cost of only little noise suppression, the non-linear spatial filter provides high noise suppression at the cost of speech distortions.

Next, we combine each spatial filter with a DNN-based post-filter that is trained independently. For this, we obtain enhancement results from the \ac{MVDR} and the NSF and use this as noisy input to the single-channel network described in Section \ref{ss:na:pf}. The results are depicted by the red bars in Figure \ref{fig:separated-processing}. We observe that independent post-filtering is far more effective when combined with a distortionless linear spatial filter than a non-linear spatial filter. Applying a tempo-spectral post-filter to the output of the MVDR nearly doubles the performance, which is clearly visible when comparing the two bottom left spectrograms in Figure \ref{fig:spectrograms}. In contrast, applying an independent post-filter to the output of the DNN-based non-linear spatial filter only slightly improves the performance by about $11$\%. This is because speech information that was lost during spatial processing cannot be recovered by multiplication with the post-filter mask.

The overall best performance is obtained by the DNN-based joint non-linear filter that does not separate the spatial and tempo-spectral processing. This filter is represented by the green bar in Figure \ref{fig:separated-processing} and it outperforms the oracle MVDR with DNN-based post-filter by 0.25 POLQA score. When comparing the spectrogram of the JNF (bottom right) with the MVDR plus post-filter (bottom left), we see that the high frequency clean speech components are preserved better. Please find audio examples on our website\footnote[1]{\footnotesize \url{https://uhh.de/inf-sp-dnn-mc-filter}}. As the joint non-linear filter clearly improves over the non-linear spatial filter plus independent post-filer, %
we conclude that the spectral and temporal information is used to enhance the non-linear spatial processing itself. Consequently, spatial processing should not be separated from tempo-spectral processing.
\section{Contribution of information sources}\label{sec:contribution}

\newcolumntype{m}[1]{D{:}{\pm}{#1}}
\begin{table}
    \caption{Impact of different sources of information included in the processing. We report mean improvements and the 95\% confidence interval.}
    \label{table:contribution}
    \centering
    \begin{tabular}{l@{}cm{-1}@{}m{-1}}\toprule
         &~~&  \multicolumn{1}{c}{$\Delta$ POLQA} & \multicolumn{1}{c}{$\Delta$ SI-SDR [dB]}\\\midrule
          PF && -0.01~:~0.01&-3.34~:~0.15\\\midrule
          F-NSF && 0.78~:~0.03 & 7.45~:~0.11\\
          T-NSF && 0.46~:~0.03 &  5.64~:~0.13\\
          FT-NSF && 0.87~:~0.03 & 7.70~:~0.12\\\midrule
          F-JNF && 1.15~:~0.04 & 8.99~:~0.12\\
          T-JNF \cite{li2019narrowband} && 0.74~:~0.03 & 7.45~:~0.13\\
          FT-JNF (proposed) && 1.43~:~0.04 & 9.94~:~0.13\\\bottomrule
    \end{tabular}
\end{table}

Next, we further investigate the contribution of different sources of information. As the dataset is very challenging with low SNR and many interfering speech sources having similar tempo-spectral structure as the target signal, spatial processing is critical for good performance. The same post-filter trained directly on the noisy input as opposed to the output of a spatial filter performs poorly as reported in the first row of Table \ref{table:contribution}. In contrast, all filters involving spatial processing provide a substantial improvement score. 

In Table \ref{table:contribution}, we compare the performance of a non-linear spatial filter with access to global spectral, temporal or tempo-spectral information. As expected, incorporating both, the temporal and spectral information, results in higher improvement scores than complementing spatial information only with one other source of information. However, the more surprising observation is that spectral information seems to be much more valuable than temporal information as suggested by the 0.32 POLQA score and $1.8$~dB SI-SDR \cite{roux2019SDRHalfbakedWell} difference in performance improvement. 

This finding does not only hold for the global information accessible to the non-linear spatial filter, but also for fine-grained information provided to the joint non-linear filter. The performance differences here amount to 0.41 POLQA score and $1.54$~dB SI-SDR. This means that our proposed slight changes to the architecture T-JNF suggested by Li and Horaud \cite{li2019narrowband}, which has originally been proposed for the CHiME3 dataset, lead to drastic performance improvements of up to 0.69 POLQA score for FT-JNF on our speech extraction dataset, which requires much stronger spatial filtering capabilities for good performance. 

\section{Conclusions}

In this paper, we have shown that non-linear spatial processing with DNNs is a key to high multi-channel speech enhancement performance. However, the potential of non-linear spatial filtering can only be fully unlocked if spatial processing is tightly integrated with tempo-spectral filtering which contradicts the traditional two-step approach of beamforming followed by post-filtering. We have furthermore shown that, in a difficult speech extraction scenario, which requires strong spatial filtering performance, spectral information is more valuable than temporal information with a difference that amounts to 0.4 POLQA score.

\bibliographystyle{IEEEtran}

\atColsBreak{\vskip5pt}
\bibliography{mybib}

\end{document}

%% file: acronyms.tex
\begin{acronym}
\acro{DFT}{discrete Fourier transform}
\acro{MVDR}{minimum variance distortionless response}
\acro{PDF}{probability density function}
\acro{MMSE}{minimum mean square error}
\acro{ML}{maximum likelihood}
\acro{SNR}{signal-to-noise ratio}
\acro{MAP}{maximum a posteriori}
\acro{ASR}{automatic speech recognition}
\acro{POLQA}{perceptual objective listening quality analysis}
\acro{MOS}{mean opinion score}
\acro{PESQ}{perceptual evaluation of speech quality}
\acro{EM}{expectation maximization}
\acro{DNN}{deep neural network}
\acro{LSTM}{long short-term memory}
\acro{FF}{feed-forward}
\acro{cIRM}{complex ideal ratio mask}
\acro{IRM}{ideal ratio mask}
\acro{STFT}{short-term Fourier transform}
\end{acronym}

%% file: notation.tex
\def\vec#1{\ensuremath{\mathbf{#1}}}

\newcommand{\vecY}{\vec{Y}}
\newcommand{\vecy}{\vec{y}}
\newcommand{\vecV}{\vec{V}}
\newcommand{\vecv}{\vec{v}}
\newcommand{\vecX}{\vec{X}}